# A Weight-Dependent 1RM Prediction Equation Optimized on 303,494 Near-Failure Sets Across 388 Exercises

Thiago Marzagao,* Ph.D.

* Author's affiliation: Fitbod, Inc.

* Author's email: thiago@fitbod.me

## PRE-PRINT

## Abstract

Classical equations for predicting one-repetition maximum (1RM) from submaximal performance — including those of Brzycki (1993), Epley (1985), Wathen (1994), and Mayhew et al. (1992) — were derived from small samples performing a single exercise (typically the bench press), yet are routinely applied to hundreds of exercises in research and practice. All such equations use a fixed conversion factor relating repetitions to estimated 1RM, regardless of the exercise or load. We used large-scale observational training data from a consumer fitness app (303,494 near-failure sets from 14,966 users across 388 exercises spanning 16 muscle groups) to derive and evaluate a generalization of the classical framework in which the conversion factor varies logarithmically with the weight lifted: 1RM = w × (1 + (r − 1)^0.85 / (−2.55 + 4.58 × ln(w))). Because the dataset contains no directly measured maxima, we optimized and evaluated the formula using an internal consistency criterion — the degree to which different weight–repetition combinations from the same person, exercise, and time window yield the same estimated 1RM. The proposed formula reduced inconsistency by 17–22% relative to all four classical benchmarks, with the improvement positive for every one of the 183 exercises with sufficient data. Five-fold user-level cross-validation confirmed near-zero overfitting: the mean test-set improvement was indistinguishable from the full-sample estimate across all benchmarks. An ablation analysis attributed 91% of the improvement to the weight-dependent conversion factor and 9% to the sub-linear repetition exponent. The conversion factor increases with load: at light weights (e.g., 10 kg dumbbell lateral raises) each additional repetition implies a larger fraction of maximal capacity than at heavy weights (e.g., 100 kg squats), consistent with prior evidence that the repetitions–%1RM relationship varies by exercise. Classical equations, by applying a single conversion factor across all loads, systematically underestimate this variation — and the discrepancy is largest for the lighter, more diverse exercises that dominate real-world training programs.

## Literature Review

### 1. Classical Prediction Equations

The estimation of one-repetition maximum (1RM) from submaximal performance has been a topic of practical and scientific interest for over four decades. Several prediction equations have been proposed, most sharing a common structure: a function relating the weight lifted and the number of repetitions performed to an estimated 1RM. Among the most widely adopted are the equations attributed to Epley (1985), Mayhew et al. (1992), and Brzycki (1993).

The Epley equation, `1RM = w × (1 + r / 30)`, originated not from a peer-reviewed study but from a poundage chart published in a resistance training manual for athletes at the University of Nebraska (Epley, 1985). The formula was later extrapolated from this chart and gained widespread adoption despite the absence of a documented empirical derivation.

Brzycki (1993) proposed the equation `1RM = w / (1.0278 − 0.0278 × r)` in a practitioner-oriented article in the *Journal of Physical Education, Recreation & Dance*. Like the Epley formula, this was not accompanied by a formal empirical study with a reported sample or methodology; it was presented as a practical tool for strength coaches.

Mayhew et al. (1992) offered a more empirically grounded contribution, testing 435 college students (184 men, 251 women) on the bench press and fitting an exponential model relating repetitions to failure to percentage of 1RM. Their equation, `%1RM = 52.2 + 41.9 × e^(−0.055 × r)`, was among the first to be derived from a substantial sample.

Other commonly cited equations from this era — including those attributed to Lander (1985), O'Conner et al. (1989), Lombardi (1989), and Wathen (1994) — similarly originated from practitioner manuals, textbook chapters (e.g., Baechle's *Essentials of Strength Training and Conditioning*), or small unpublished datasets, rather than from peer-reviewed empirical studies with large samples.

A common feature of all classical equations is that they were derived from a narrow empirical base: typically young, trained males performing the bench press. Whether and how well these equations generalize to other exercises, populations, and training contexts has been a central concern in subsequent research.

## 2. Validation and Comparison Studies

Several studies have directly compared the accuracy of classical prediction equations against measured 1RM values.

LeSuer et al. (1997) tested seven equations on 67 untrained college students (40 men, 27 women) across the bench press, squat, and deadlift. While correlations between predicted and actual 1RM were uniformly high (r > 0.95), there were meaningful differences in absolute accuracy across exercises. In particular, all equations significantly underestimated the deadlift 1RM, providing early evidence that a single equation may not generalize across exercises.

Reynolds et al. (2006) examined 70 participants (34 men, 36 women) aged 18–69 on the bench press and leg press, comparing predictions from 1RM, 5RM, 10RM, and 20RM tests. The 5RM test produced the highest prediction accuracy ($R^2$ = 0.993 for bench press, $R^2$ = 0.974 for leg press), with accuracy degrading substantially at higher repetition ranges. Notably, the inclusion of anthropometric variables did not improve predictions beyond the submaximal load itself.

Mayhew et al. (2008) evaluated prediction accuracy in 103 women before and after 12 weeks of resistance training, finding that equations were more accurate when fewer than 10 repetitions to failure were used. Training status (pre- vs. post-training) affected accuracy, though the direction of this effect was exercise- and equation-dependent.

Across these and other validation studies, several consistent findings emerge. First, all classical equations produce high correlations with actual 1RM (typically r > 0.90), which can obscure meaningful absolute errors. Second, prediction accuracy degrades at higher repetition ranges (above approximately 10 repetitions). Third, accuracy varies across exercises, with the deadlift and lower-body compound movements often showing larger prediction errors than the bench press.

## 3. Exercise-Specificity of the Repetitions–%1RM Relationship

A body of evidence indicates that the relationship between repetitions to failure and percentage of 1RM is not uniform across exercises — a finding with direct implications for the validity of universal prediction equations.

Shimano et al. (2006) tested trained and untrained men on the back squat, bench press, and arm curl at 60%, 80%, and 90% of 1RM. At 60% 1RM, participants completed significantly more repetitions on the back squat than on the bench press or arm curl, suggesting that the volume of muscle mass engaged modulates repetition capacity at a given relative intensity. Training status had minimal impact on this relationship.

Richens and Cleather (2014) compared 8 competitive weightlifters and 8 endurance runners on the leg press at 70%, 80%, and 90% of 1RM. Endurance runners completed dramatically more repetitions than weightlifters at 70% (39.9 vs. 17.9) and 80% (19.8 vs. 11.8) of 1RM, with the difference narrowing at 90%. This study demonstrates that not only exercise type but also training background can fundamentally alter the repetitions–%1RM curve.

The most comprehensive analysis to date is the meta-regression by Nuzzo et al. (2024), which synthesized 952 repetition-to-failure tests from 7,289 individuals across 269 studies. Exercise type was the only moderator that meaningfully affected the repetitions–%1RM relationship; sex, age, and training status had minimal influence. The relationship was best described by natural cubic splines rather than the linear or simple exponential forms used in classical equations. Separate loading tables were provided for the bench press and leg press, as these exercises showed distinctly different curves.

Despite this large meta-analysis, the literature remains heavily skewed toward a small number of exercises: bench press accounted for 42% of included tests and leg press for 14%. Dozens of commonly performed exercises — including rows, overhead presses, deadlift variations, and isolation movements for smaller muscle groups — lack validated exercise-specific equations.

## 4. Fatigue, Set Order, and Rest Interval Effects

Classical 1RM prediction equations treat each set as an independent observation, implicitly assuming that performance on any given set reflects maximal capacity at the prescribed load. In practice, however, performance declines across successive sets due to neuromuscular fatigue, and this decline is modulated by rest interval duration.

Willardson (2006) reviewed the factors affecting rest intervals between resistance exercise sets and concluded that 3–5 minutes of rest is necessary to maintain training intensity for strength-oriented goals, while shorter intervals (1–2 minutes) produce meaningful declines in repetition performance.

Senna et al. (2011) directly quantified this effect in 15 trained males performing multi-joint (bench press, leg press) and single-joint (chest fly, leg extension) exercises with 1-, 3-, or 5-minute rest intervals. Significant repetition count declines were observed starting from the second set with 1-minute rest and from the third set with 3- and 5-minute rest. Perceived exertion increased progressively across sets for all rest conditions, with significantly higher values under shorter rest.

These findings have implications for 1RM estimation from training log data: a set performed third in a sequence with 90 seconds of rest carries different information about maximal capacity than a first set performed with full recovery. No existing 1RM prediction equation accounts for set order or accumulated within-session fatigue.

## 5. Proximity to Failure and Reps in Reserve

A fundamental assumption underlying all repetition-based 1RM prediction is that the set is performed to, or near, momentary muscular failure. In practice — particularly in unsupervised recreational training — this assumption is frequently violated.

Zourdos et al. (2016) introduced a resistance training-specific rating of perceived exertion (RPE) scale based on repetitions in reserve (RIR), in which RPE 10 corresponds to 0 RIR (failure). Testing 29 squatters (15 experienced, 14 novice), they found strong inverse correlations between movement velocity and RPE for both experienced ($r = -0.88$) and novice ($r = -0.77$) participants, supporting the scale's construct validity.

Steele et al. (2017) examined 141 participants' ability to predict repetitions to momentary failure and found systematic underprediction across experience levels: experienced trainees underpredicted by approximately 1–2 repetitions, while less experienced trainees underpredicted by approximately 4–5 repetitions. This indicates that even when trainees believe they are at or near failure, they typically have multiple repetitions in reserve.

Refalo et al. (2024) evaluated intraset RIR prediction accuracy in 24 resistance-trained individuals (12 men, 12 women) during the bench press at 75% 1RM. Trained subjects showed high accuracy (mean error: 0.65 ± 0.78 repetitions), with no meaningful effect of sex, training experience, or relative strength.

The implications for 1RM estimation from observational training data are substantial. If recreational app users typically train with 3–5 reps in reserve (as the Steele et al. findings suggest), then applying standard prediction equations to their logged performances will systematically underestimate true 1RM. Any data-driven approach to 1RM estimation from training logs must contend with this effort confound.

## 6. Velocity-Based Approaches

An alternative paradigm for 1RM estimation uses movement velocity rather than repetition counts. González-Badillo and Sánchez-Medina (2010) demonstrated in 120 strength-trained males that mean propulsive velocity during the bench press was closely related to %1RM ($R^2 = 0.98$). Importantly, this relationship remained stable even after a 9.3% increase in 1RM following six weeks of training, meaning velocity tracks relative intensity regardless of changes in absolute strength.

A systematic review and individual participant data meta-analysis by Greig et al. (2023) pooled 434 participants across 20 studies and found that individualized load-velocity profiles yielded a pooled standard error of estimate of approximately 9.8% of 1RM. While these findings support the utility of velocity-based approaches for monitoring, the authors recommended direct 1RM assessment when precision is required.

The velocity-based literature, although not directly applicable to contexts where bar speed is not measured (such as most consumer fitness apps), establishes an important principle: individualized models consistently outperform generic equations. This insight transfers to repetition-based estimation as well.

## 7. Large-Scale and Data-Driven Approaches

Despite the proliferation of prediction equations and their validation studies, the literature remains characterized by small sample sizes and narrow exercise coverage. The largest empirical studies (e.g., Mayhew et al., 1992, N = 435; Reynolds et al., 2006, N = 70) are modest by the standards of contemporary data science. Even the Nuzzo et al. (2024) meta-analysis, which aggregated across 269 studies, encompassed only 7,289 individuals — a fraction of the data generated by a single month of usage on a modern resistance training app.

The OpenPowerlifting project maintains an open archive of over 2 million competition results from competitive powerlifters, which has been used for research on scoring formulas (e.g., Wilks vs. IPF). However, this dataset contains directly tested competition maxima rather than submaximal training data, and is limited to three exercises (squat, bench press, deadlift), making it unsuitable for deriving or validating general 1RM prediction equations.

To date, no published study has used large-scale observational training log data to derive or validate 1RM prediction equations. This represents a significant gap, both methodological and practical.

## 8. Summary of Knowledge Gaps

The preceding review identifies several unresolved issues:

1. **Narrow empirical basis for classical equations.** The most widely used prediction equations were derived from small samples, limited to one or two exercises, and in some cases lack formal empirical documentation entirely.
2. **Exercise-specificity without exercise-specific equations.** It is well established that the repetitions–%1RM relationship varies by exercise, yet validated exercise-specific (or exercise-category-specific) prediction equations exist only for the bench press and, to a lesser extent, the leg press and squat.
3. **No accounting for within-session fatigue.** All existing equations treat each set as independent, ignoring the well-documented effect of set order and rest interval on repetition performance.
4. **Absence of large-scale, data-driven derivations.** The methodological tools and datasets now exist to derive prediction equations from hundreds of thousands or millions of training observations, yet no such study has been conducted.
5. **No alternative optimization criterion for observational data.** Existing equations were fit (or intuited) to minimize error against directly tested 1RM. When ground-truth 1RM data are unavailable — as in observational training logs — alternative criteria for equation derivation are needed but have not been proposed.

The present study addresses gaps 2, 4, and 5 by leveraging a large-scale dataset of logged resistance training sessions to derive exercise-category-specific prediction equations using an internal consistency optimization criterion.

# Methods

## Data Source

The data for this study were drawn from Fitbod, a resistance training app. When users complete a workout — whether generated by the app's recommendation algorithm or created manually — the app records each set they perform: the exercise, the weight used, the number of repetitions completed, and whether the set was flagged as AMRAP (an "As Many Reps As Possible" set, in which the user intends to continue until failure). This provides a large-scale observational record of real-world resistance training behavior.

We extracted a 5% deterministic random sample of users from Fitbod's database (sampled by hashing user identifiers, ensuring reproducibility). The full user base would yield substantially more data, but a 5% sample was chosen for computational tractability: the optimization and cross-validation procedures described below involve evaluating millions of formula candidates across hundreds of thousands of sets, and preliminary tests indicated that increasing the sample size beyond 5% produced negligible changes in the optimized coefficients. The raw extract contained 37,736,594 sets from 65,757 users across 505 exercises.

A note on weight logging conventions: in Fitbod, users log the weight per hand for dumbbell exercises (e.g., a user pressing two 25 kg dumbbells logs "25 kg") and the total barbell weight (including the bar) for barbell exercises. For machine and cable exercises, the logged weight is the weight stack setting. These conventions are consistent within and across users by design of the app's interface, but they mean that the absolute weight values reflect different things for different equipment types. This does not affect within-exercise comparisons — all tuples in our analysis compare sets from the same user on the same exercise — but it does mean that the weight-dependent k(w) function should be interpreted as capturing the statistical relationship between logged weight and the rep–1RM curve, not as a direct measure of total external load.

## Data Processing Pipeline

Starting from the raw extract, we applied a series of filters in sequence. The goal was to arrive at a sample of sets that are genuinely informative for 1RM estimation — that is, sets where the user was likely exerting maximal effort at the given load, performed on exercises where the logged weight unambiguously represents the external resistance.

**Step 1. Extraction-level filters (applied at query time).** The following were excluded during data extraction:

- Warmup sets (tagged by the app; these are deliberately sub-maximal)
- Sets with zero weight or with more than 30 repetitions
- Custom exercises (user-created exercises that do not appear in Fitbod's standardized exercise library; because they lack curated metadata — muscle group, compound vs. isolation classification, etc. — they cannot be reliably categorized or compared across users)
- Non-resistance exercises (cardio, mobility, and timed exercises, which do not involve the kind of load-repetition tradeoff that 1RM equations describe)
- Bodyweight and assisted exercises (e.g., pull-ups, dips, assisted chin-ups; for these exercises, the recorded weight reflects only the *added* weight, not the total resistance the user is working against, making 1RM estimation ambiguous)
- Users without a recorded gender or date of birth

**Step 2. Ethics filter (minors).** All users who appeared anywhere in the data as under 18 years of age were excluded entirely — not just the individual sets recorded before age 18, but *all* of that user's data regardless of when it was logged. This conservative approach was adopted on ethical grounds: even though many of these users turned 18 at some point during their training history, retroactive inclusion would still involve data collected from minors. This removed 4,185 users (leaving 61,572).

**Step 3. Quality filters.** We removed exercises that lacked a compound/isolation classification in Fitbod's exercise metadata, sets with weight ≥ 500 kg (apparent data entry errors), and exercises containing "Ball" in their name (stability ball exercises where the logged "weight" may refer to the ball itself rather than external resistance). After this step: 36,632,918 sets from 61,546 users across 450 exercises.

**Step 4. Near-failure set selection.** A fundamental challenge with observational training data is that most sets are *not* performed to failure. If a user performs 8 reps of bench press at 80 kg but could have done 12, the set carries little information about their true 1RM — the mapping from (weight, reps) to 1RM depends on the set being a maximal or near-maximal effort at that load. We addressed this by restricting the sample to sets showing evidence of near-failure effort, using two complementary signals:

1. *AMRAP sets.* The user explicitly flagged the set as "As Many Reps As Possible," indicating an intent to continue until failure. Across the quality-filtered data, 8.2% of sets carried this flag.
2. *Fatigue-detected sets.* On the same exercise within the same workout, the user performed fewer reps than on a preceding set at the same weight. For example, if a user does 10 reps at 60 kg and then 8 reps at 60 kg, the second set shows a fatigue-induced performance decline — evidence that the user was at or near their limit. This captures near-failure effort even when the user did not use the AMRAP flag.

Together, these two criteria identified 5,316,974 near-failure sets (approximately 50% AMRAP-only, 43% fatigue-detected only, and 7% meeting both criteria).

**Step 5. First-set-per-workout deduplication.** To avoid double-counting the cumulative effects of within-workout fatigue, we retained only the *first* near-failure set per user per exercise per workout day. This keeps the freshest data point — the one least contaminated by accumulated fatigue from earlier sets. After this step: 3,391,418 sets.

**Step 6. Time windows and observation tuples.** Each set was assigned to a 14-day calendar window (fixed buckets anchored to January 1, 2000). We then defined an *observation tuple* as a unique (user, exercise, 14-day window) combination. A tuple contains all the near-failure sets that a given user performed on a given exercise within a given two-week period. The choice of 14 days balances two competing concerns: windows must be short enough that a user's true strength is approximately stable within a window, but long enough to accumulate multiple sets at varying rep ranges for the same exercise.

For a tuple to be informative, it must contain variation in both inputs to a prediction formula. We required each tuple to contain at least 2 distinct rep counts *and* at least 2 distinct weights. Tuples with less variation (e.g., a user who always benches 80 kg for 8 reps) provide no signal for distinguishing between prediction formulas.

This final filter produced the analysis sample: 303,494 sets in 135,730 tuples from 14,966 users across 388 exercises.

Table 1 summarizes the complete pipeline.

**Table 1. Data processing pipeline.**

| Step | Description | Sets | Users |
| --- | --- | --- | --- |
| 0 | Raw extract (5% sample) | 37,736,594 | 65,757 |
| 1 | Extraction-level filters | (applied at query time) | — |
| 2 | Remove users ever <18 | — | 61,572 |
| 3 | Quality filters | 36,632,918 | 61,546 |
| 4 | Near-failure selection | 5,316,974 | — |
| 5 | First-per-workout dedup | 3,391,418 | — |
| 6 | Tuple qualification | 303,494 | 14,966 |

## Participants

Table 2 summarizes the demographic characteristics of the 14,966 users in the final analysis sample.

**Table 2. Participant demographics (N = 14,966).**

|  | n or value |
| --- | --- |
| **Gender** | |
| Male | 11,964 (79.9%) |
| Female | 3,002 (20.1%) |
| **Age (years)** | |
| Mean (SD) | 35.0 (9.7) |
| Median (IQR) | 34 (28–41) |
| Range | 18–82 |
| **Age distribution** | |
| 18–24 | 1,949 (13.0%) |
| 25–29 | 2,930 (19.6%) |
| 30–34 | 3,129 (20.9%) |
| 35–39 | 2,619 (17.5%) |
| 40–44 | 1,878 (12.6%) |
| 45–49 | 1,209 (8.1%) |
| 50–54 | 667 (4.5%) |
| 55–59 | 344 (2.3%) |
| 60+ | 239 (1.6%) |

*Note.* Age was computed as the year-level difference between workout date and date of birth; age statistics are based on each user's median age across their records. Two users with implausible ages (>100 years) were retained in the analysis but excluded from this table (N = 14,964 for age statistics).

The sample is predominantly male and concentrated in the 25–39 age range, reflecting the user base of a consumer fitness app rather than a representative population sample. Unlike laboratory studies, which typically recruit homogeneous convenience samples (e.g., college-age males performing the bench press), this dataset captures a broad range of training experience, body types, and exercise preferences in naturalistic training conditions.

## Analysis Sample

Table 3 summarizes the characteristics of the final analysis sample.

**Table 3. Analysis sample characteristics.**

|  | Value |
|---|---|
| **Sample size** | |
| Sets | 303,494 |
| Observation tuples | 135,730 |
| Users | 14,966 |
| Exercises | 388 |
| **Exercise type** | |
| Compound | 222 exercises (68,937 tuples) |
| Isolation | 166 exercises (66,793 tuples) |
| **Repetitions** | |
| Mean (SD) | 9.0 (3.5) |
| Median | 8 |
| Range | 1–30 |
| **Weight (kg)** | |
| Mean (SD) | 41.3 (32.3) |
| Median (IQR) | 31.8 (18.1–59.0) |
| **Sets per tuple** | |
| Mean | 2.2 |
| Median | 2 |
| Tuples with exactly 2 sets | 81.3% |
| Tuples with ≥3 sets | 18.7% |
| **Tuples per user** | |
| Mean | 9.1 |
| Median | 3 |

The dataset covers 16 primary muscle groups. Table 4 shows the breakdown by muscle group.

**Table 4. Analysis sample breakdown by primary muscle group.**

| Muscle group | Exercises | Tuples | Sets | Mean weight (kg) |
|---|---|---|---|---|
| Chest | 43 | 35,400 | 79,574 | 46.3 |
| Biceps | 46 | 19,532 | 44,030 | 20.3 |
| Triceps | 21 | 17,030 | 37,985 | 29.2 |
| Back | 33 | 16,735 | 36,656 | 47.8 |
| Shoulders | 59 | 15,941 | 35,836 | 22.6 |
| Quadriceps | 67 | 9,909 | 21,744 | 76.2 |
| Hamstrings | 42 | 7,280 | 15,907 | 54.4 |
| Abs | 21 | 3,746 | 9,069 | 46.1 |
| Calves | 14 | 2,488 | 5,622 | 80.3 |
| Trapezius | 14 | 2,337 | 5,152 | 44.9 |
| Glutes | 11 | 1,595 | 3,493 | 58.8 |
| Abductors | 3 | 1,051 | 2,358 | 67.8 |
| Adductors | 2 | 1,044 | 2,330 | 63.9 |
| Forearms | 7 | 811 | 1,894 | 18.9 |

| Muscle group | Exercises | Tuples | Sets | Mean weight (kg) |
|---|---|---|---|---|
| Lower Back | 4 | 745 | 1,647 | 68.1 |
| Neck | 1 | 86 | 197 | 7.9 |

Chest is the most represented muscle group, driven largely by the Barbell Bench Press (9,850 tuples), followed by Biceps, Triceps, Back, and Shoulders. Lower-body muscle groups are less represented (Quadriceps: 9,909 tuples; Hamstrings: 7,280; Calves: 2,488), reflecting typical usage patterns in a general-population fitness app. The five most common individual exercises were the Barbell Bench Press, Lat Pulldown, Dumbbell Bench Press, Dumbbell Bicep Curl, and Cable Rope Tricep Extension. Mean weight lifted ranged from 7.9 kg (Neck) to 80.3 kg (Calves), reflecting the wide variety of exercises and loading patterns in the dataset.

## Optimization Criterion

Traditional approaches to deriving 1RM prediction equations involve collecting a ground-truth 1RM measurement (via a direct maximal test) and fitting a formula to minimize the discrepancy between predicted and observed 1RM. This approach is not available here: the data consist entirely of submaximal training logs with no directly tested 1RM values.

We instead adopt an **internal consistency** criterion. The key insight is straightforward: if a prediction formula is correct, then different sets from the same person on the same exercise within a short time period should all produce the *same* estimated 1RM. A user's true bench press max does not change meaningfully between Tuesday and the following Thursday. So if on Tuesday they bench 80 kg for 8 reps, and on Thursday they bench 90 kg for 5 reps, a good formula should map both observations to approximately the same 1RM estimate. A bad formula will produce scattered estimates.

Formally, we measure consistency as the **mean within-tuple standard deviation of log-transformed 1RM estimates**, denoted SD(log(1RM)), computed across all 135,730 tuples in the dataset. The formula that minimizes this metric is the one that produces the most internally consistent 1RM estimates across different weight–rep combinations for the same user on the same exercise.

We use the log transformation for two reasons. First, 1RM values span a wide range (from under 10 kg for a Dumbbell Lateral Raise to over 200 kg for a Deadlift), and using raw values would let heavy exercises dominate the metric. Log-transforming makes the metric scale-invariant: a 5 kg discrepancy matters more for a 20 kg lift than for a 200 kg lift, which is the desirable behavior. Second, the log transformation prevents a mathematical artifact that could otherwise distort the optimization. Without the log, a formula could artificially reduce the raw standard deviation by systematically inflating estimates — stretching them upward in a way that compresses the SD relative to the mean, without actually improving consistency. Operating in log-space makes the metric immune to this kind of multiplicative scaling: a formula cannot game its score by systematically inflating or deflating estimates.

## Formula Structure

Classical 1RM prediction equations share a common structure. The Epley equation, `1RM = w × (1 + r / 30)`, adds a fixed fraction of the weight for each additional repetition: do one more rep, add w/30 to the estimate. The Brzycki equation, `1RM = w / (1.0278 − 0.0278 × r)`, is algebraically similar at low rep counts but diverges at higher reps (it sends 1RM to infinity as reps approach 36, while Epley grows linearly without bound).

Both equations implicitly assume that the relationship between reps and %1RM is the same regardless of absolute load. That is, going from 5 reps to 10 reps corresponds to the same relative change in estimated 1RM whether the user is curling 10 kg or squatting 150 kg. This assumption has been questioned on theoretical grounds (larger muscle groups may sustain more repetitions at a given %1RM; see Shimano et al., 2006), but no published equation to date has modeled the dependency explicitly.

Our formula generalizes the Epley structure in two ways.

**Weight-dependent rep-to-1RM conversion.** Instead of a fixed conversion factor (the constant 30 in Epley, or effectively 1/0.0278 ≈ 36 in Brzycki), we allow the conversion factor to depend on the weight lifted:

```
k(w) = a + b × ln(w/w₀)
```

where `w` is the weight in kilograms, `w₀ = 1 kg` is a reference weight (included to make the logarithm's argument dimensionless), `ln` is the natural logarithm, and `a` and `b` are parameters to be estimated from the data. Because `w₀ = 1 kg`, the expression simplifies algebraically to `a + b × ln(w)` when `w` is expressed in kilograms, but the `w/w₀` formulation makes explicit that the argument is a dimensionless ratio. (In imperial units, with `w₀ = 1 lb`, the optimized intercept `a` would shift by `b × ln(2.205)` while `b` itself and all model predictions would remain unchanged.)

Conceptually, k(w) controls how steeply each additional repetition increases the 1RM estimate. A smaller k means each rep is "worth more" in 1RM terms; a larger k means each rep contributes less. The logarithmic form allows k to increase with weight at a decelerating rate — the difference between 10 kg and 20 kg matters more than the difference between 100 kg and 110 kg.

We hypothesize that this pattern reflects a physiological regularity: lighter exercises tend to involve smaller, more fatigue-susceptible muscle groups, where each additional rep represents a larger fraction of the remaining capacity, while heavier compound lifts recruit larger muscle groups that can sustain more reps at a given percentage of their maximum. This hypothesis is consistent with the exercise-specificity findings of Shimano et al. (2006) and the meta-regression of Nuzzo et al. (2024). However, we note that weight in our dataset serves as a statistical proxy for exercise type, equipment category, and muscle group — not as a direct measure of physiological load. Two exercises logged at the same weight (e.g., a 25 kg dumbbell bench press and a 25 kg concentration curl) may have genuinely different rep–%1RM curves that the formula cannot distinguish. The extent to which weight-dependence captures true physiological modulation versus equipment-specific logging conventions is an empirical question that future work should address.

**Sub-linear repetition scaling.** Instead of using raw `(r − 1)` as in Epley, we raise the rep-minus-one term to a power α:

```
(r − 1)^α
```

where α is a parameter between 0 and 1. When α < 1, this introduces diminishing returns: the jump from 1 rep to 2 reps contributes more to the 1RM estimate than the jump from 14 reps to 15 reps. This reflects the observation that at higher rep counts, factors beyond pure muscular strength — cardiovascular endurance, pain tolerance, mental fatigue — increasingly influence when the user stops.

The complete formula is:

```
1RM = w × (1 + (r − 1)^α / k(w))
    = w × (1 + (r − 1)^α / (a + b × ln(w)))
```

where w is the weight lifted (kg) and r is the number of repetitions performed. The formula has three parameters: α, a, and b.

Note that for r = 1 the formula reduces to 1RM = w, as expected: if a user lifts a weight for exactly one rep, the best estimate of their 1RM is that weight itself. For r > 1, the formula adds a correction that depends on both the number of reps and the absolute weight.

A computational guard k(w) ≥ 0.5 is imposed to prevent division by near-zero values. As reported in the Robustness Analyses section, this guard activates for only 173 of 303,494 sets (0.06%), all involving weights below 2 kg, confirming that it is effectively inert for the analysis.

## Optimization Procedure

Based on preliminary exploration, α was fixed at 0.85 prior to the main optimization. The sensitivity of results to this choice is examined in the Robustness Analyses section, which shows that the consistency improvement is positive across the full range α ∈ [0.50, 1.00] and that the weight-dependent k(w) — not the choice of α — drives the bulk of the improvement.

With α fixed, the remaining two parameters (a, b) were optimized via a two-stage grid search to minimize the mean within-tuple SD(log(1RM)) across all tuples in the dataset:

**Coarse pass.** We evaluated 9,800 (a, b) combinations: a ranged from −30.0 to 19.5 in steps of 0.5 (100 values), and b ranged from 0.5 to 19.9 in steps of 0.2 (98 values). For each combination, we computed the estimated 1RM for every set in the sample and then calculated the mean within-tuple SD(log(1RM)). A floor of k(w) ≥ 0.5 was imposed to prevent division by near-zero values.

**Refinement pass.** We then searched a fine grid centered on the coarse optimum: a ± 0.5 in steps of 0.05 (21 values) and b ± 0.2 in steps of 0.02 (21 values), totaling 441 combinations. The (a, b) pair that minimized SD(log(1RM)) in this refinement pass was taken as the final estimate.

## Cross-Validation

To assess whether the optimized coefficients generalize beyond the specific sample used for fitting, we performed 5-fold cross-validation with a critical design choice: **splits were made at the user level, not the set level**.

This means that all data from a given user — every exercise, every tuple, every set — appeared in either the training fold or the test fold, but never both. This is the most stringent form of cross-validation available for this data: the formula must generalize to *entirely new people*, not merely to new sets from users it has already seen. Set-level splitting would be substantially more lenient (and would flatter the results), because the model could exploit within-user patterns already encountered during training.

For each of the five folds:

1. Users were randomly assigned (with a fixed seed for reproducibility) to an 80% training set and a 20% test set.
2. The full two-stage grid search was re-run from scratch on the training tuples, producing fold-specific values of a and b.
3. The fold-specific coefficients were applied to the held-out test tuples to compute SD(log(1RM)).
4. The Brzycki equation was also evaluated on the same test tuples, providing a within-fold benchmark.

Overfitting was quantified as the percentage difference between test and training SD(log(1RM)): a value near zero indicates that performance on held-out users matches performance on the training sample.

## Per-Exercise Analysis

To assess whether the formula's improvement over Brzycki is uniform across exercises or driven by a handful of them, we computed the consistency improvement separately for each exercise with at least 50 tuples. For each qualifying exercise, the Brzycki SD(log(1RM)) and the proposed formula's SD(log(1RM)) were computed and compared. This per-exercise analysis used the full-sample optimized coefficients (not the fold-specific coefficients from cross-validation).

## Benchmarks

We compare the proposed formula against the four classical equations most frequently included in published meta-analyses comparing 1RM prediction methods (LeSuer et al., 1997; Reynolds et al., 2006; Nuzzo et al., 2024):

1. **Brzycki (1993):** `1RM = w / (1.0278 − 0.0278 × r)`
2. **Epley (1985):** `1RM = w × (1 + r / 30)`
3. **Wathen (1994):** `1RM = 100 × w / (48.8 + 53.8 × e^(−0.075 × r))`
4. **Mayhew et al. (1992):** `1RM = 100 × w / (52.2 + 41.9 × e^(−0.055 × r))`

The improvement metric used throughout this paper is the percentage reduction in mean within-tuple SD(log(1RM)):

```
Improvement = (SD_classical − SD_ours) / SD_classical × 100%
```

A positive value indicates that the proposed formula produces more internally consistent 1RM estimates than the classical benchmark.

---

# Results

## Optimized Coefficients and the Optimization Surface

The two-stage grid search yielded the following optimized parameters: a = −2.55, b = 4.58, with α fixed at 0.85. The resulting formula is:

```
1RM = w × (1 + (r − 1)^0.85 / (−2.55 + 4.58 × ln(w)))
```

Applied to the full analysis sample (303,494 sets across 135,730 tuples), this formula produced a mean within-tuple SD(log(1RM)) of 0.0847. Table 5 compares this to four classical benchmarks.

**Table 5. Full-sample consistency comparison.**

| Formula | SD(log(1RM)) | Improvement |
|---|---|---|
| **Proposed** | **0.0847** | — |
| Brzycki (1993) | 0.1028 | +17.6% |
| Epley (1985) | 0.1026 | +17.4% |
| Wathen (1994) | 0.1021 | +17.0% |
| Mayhew (1992) | 0.1084 | +21.9% |

*Note.* Improvement = reduction in SD(log(1RM)) relative to the proposed formula: (classical − proposed) / classical × 100%.

The proposed formula reduces within-tuple inconsistency by 17–22% relative to every classical benchmark. The four classical equations, despite their different functional forms (linear, reciprocal, and exponential), produce similar consistency levels — the three closest (Brzycki, Epley, Wathen) span a range of only 0.0007, while Mayhew's exponential form performs slightly worse (SD = 0.1084). This suggests that the classical equations share a common structural limitation rather than differing in calibration. As we show below, that limitation is the use of a fixed conversion factor across all exercises and weight levels.

Figure 1 shows the objective function — mean within-tuple SD(log(1RM)) — evaluated across a broad range of (a, b) values. The surface reveals a well-defined valley of near-optimal solutions oriented along a diagonal, reflecting the fact that many combinations of a and b can produce similar k(w) values at moderate weights. However, the global minimum is clearly localized, and the coefficient values found by the grid search (a = −2.55, b = 4.58, marked with a star) sit squarely at the bottom of this valley. The dark regions in the lower-left corner correspond to (a, b) combinations that produce negative or near-zero k values for some weights, causing the formula to produce extreme 1RM estimates and high inconsistency.

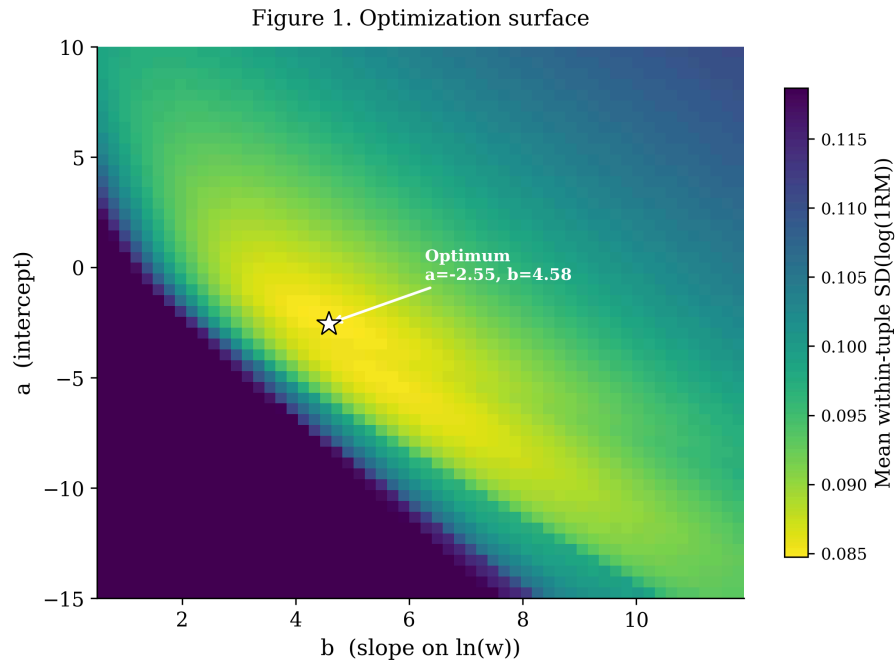

Figure 1. Optimization surface

## The Weight-Dependent Conversion Factor

The negative intercept (a = −2.55) and positive slope (b = 4.58) of the k(w) function mean that the conversion factor increases logarithmically with weight. Table 6 illustrates the implied k values at representative weight levels.

**Table 6. Implied k(w) at representative weight levels.**

| Weight (kg) | Example exercise | k(w) | Interpretation |
| --- | --- | --- | --- |
| 10 | Dumbbell Lateral Raise | 8.0 | Each rep is "worth" a large % of 1RM |
| 15 | Dumbbell Bicep Curl | 9.9 | Steep rep-to-1RM curve |
| 25 | Dumbbell Bench Press | 12.2 | Moderately steep |
| 55 | Lat Pulldown | 15.8 | Moderate |
| 70 | Barbell Bench Press | 16.9 | Moderate |
| 80 | Back Squat | 17.5 | Moderate-to-shallow |
| 150 | Heavy Deadlift | 20.4 | Shallow (each rep adds little to 1RM est.) |

For context, the Epley equation uses a fixed k = 30, the Wathen equation an effective k ≈ 29 (varying slightly with rep count), the Mayhew equation an effective k ≈ 29 (also varying with rep count), and the Brzycki equation an effective k ≈ 36. All four are far higher than the values observed at any weight in this dataset. This implies that classical equations systematically underestimate the contribution of each additional repetition to the 1RM estimate — and that this underestimation is most severe for lighter exercises, where the optimized k is as low as 8.

Figure 2 plots the k(w) curve with the mean weights of six common exercises marked along it. The horizontal reference lines mark the fixed k values of the four classical equations; the green shaded band indicates the range of Wathen's effective k across rep counts 5–10, and the purple shaded band indicates the range of Mayhew's effective k across the same range.

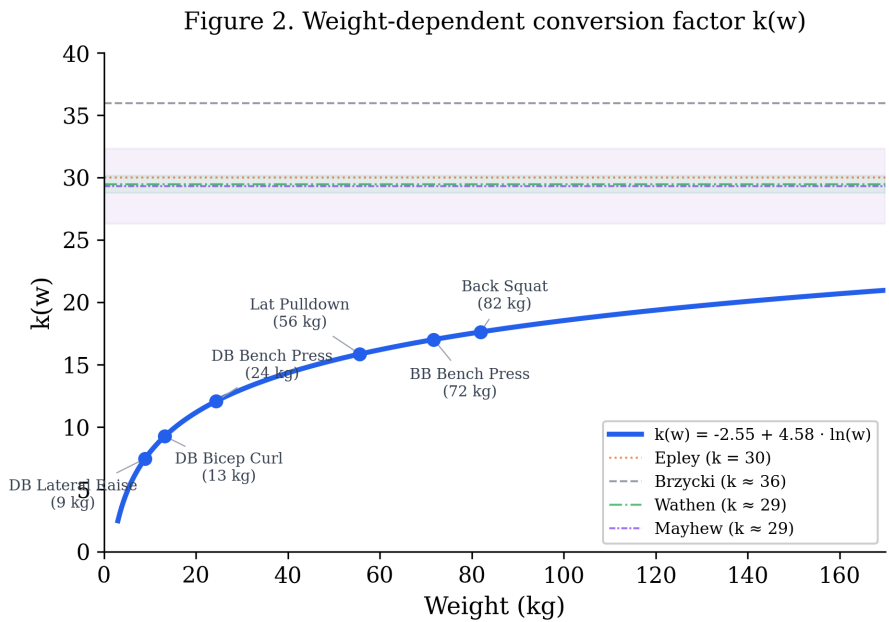

Figure 2. Weight-dependent conversion factor k(w)

The curve's shape is consistent with a hypothesis of exercise-type modulation: lighter exercises — which typically involve smaller, more fatigue-susceptible muscle groups — show lower k values, implying that each additional repetition represents a larger fraction of the remaining capacity. At heavier loads — compound movements recruiting large muscle masses — k increases, and the rep-to-%1RM curve flattens. This pattern is consistent with the exercise-specificity findings of Shimano et al. (2006) and the meta-regression of Nuzzo et al. (2024), which identified exercise type as the dominant moderator of the repetitions–%1RM relationship.

## How the Formulas Diverge in Practice

To build intuition for what the 17–22% improvement means in practice, Figure 3 compares the 1RM predictions of the four classical equations and the proposed formula across three weight levels: light (12 kg), medium (45 kg), and heavy (100 kg).

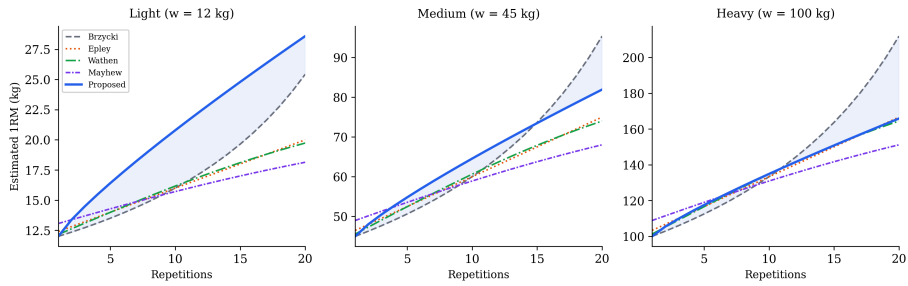

Figure 3. Predicted 1RM by formula and weight level

At light weights, the formulas diverge substantially: for 12 kg at 15 reps, the classical equations cluster around 19–20 kg, while the proposed formula predicts roughly 25 kg. The proposed formula's steeper curve at low weights (k(12) ≈ 8.8, far below Brzycki's k ≈ 36, Epley's k = 30, Wathen's k ≈ 29, or Mayhew's k ≈ 29) reflects its key insight: each additional rep carries more information about maximal capacity when the load is light.

As a concrete example, consider a 13 kg dumbbell curl at 10 reps. The classical Brzycki equation estimates a 1RM of about 17 kg; the proposed formula estimates about 22 kg. At a 100 kg bench press at 5 reps, the two formulas agree within 2 kg.

At heavy weights (100 kg), all five curves are much closer, though the proposed formula remains slightly steeper at higher rep counts. This convergence is expected: as weight increases, k(w) rises toward the range of the classical equations, and the formulas agree more closely.

The shaded area between the Brzycki and proposed curves represents the region of greatest disagreement. This region is largest for light weights and high reps — exactly the conditions under which classical equations are known to be least accurate (Reynolds et al., 2006; Mayhew et al., 2008).

### Per-Exercise Improvement

The consistency improvement was positive for **all 183 exercises** with at least 50 tuples — not a single exercise showed higher inconsistency under the proposed formula than under any of the four classical benchmarks. Table 7 breaks the results down by exercise category.

**Table 7. Per-exercise consistency improvement by category.**

| Category | n exercises | vs. Brzycki | vs. Wathen | vs. Epley | vs. Mayhew | Mean weight (kg) |
| --- | --- | --- | --- | --- | --- | --- |
| All (≥50 tuples) | 183 | +19.3% | +19.4% | +19.7% | +23.9% | 35.8 |
| Compound | 86 | +16.3% | +16.9% | +17.3% | +21.7% | 43.4 |
| Isolation | 97 | +21.9% | +21.6% | +21.8% | +25.9% | 29.2 |

The larger improvement for isolation exercises is consistent with the formula's design: isolation movements typically use lighter weights, where the weight-dependent k departs most from the classical equations' fixed conversion factor. The correlation between mean exercise weight and consistency improvement was r = −0.61 — a strong negative relationship confirming that lighter exercises benefit more.

Figure 4 makes this relationship visible: each point represents one exercise, plotted by mean weight (x-axis) and improvement over Brzycki (y-axis). The downward trend is clear, with the lightest exercises (dumbbell lateral raises, dumbbell curls) clustering in the upper left and the heaviest (barbell squats, trap bar deadlifts) in the lower right. Compound exercises (blue) and isolation exercises (orange) overlap substantially but with isolation exercises shifted toward lighter weights and larger improvements. The pattern is virtually identical when plotted against Wathen, Epley, or Mayhew (the four classical SDs differ by less than 6%), so only the Brzycki comparison is shown for clarity.

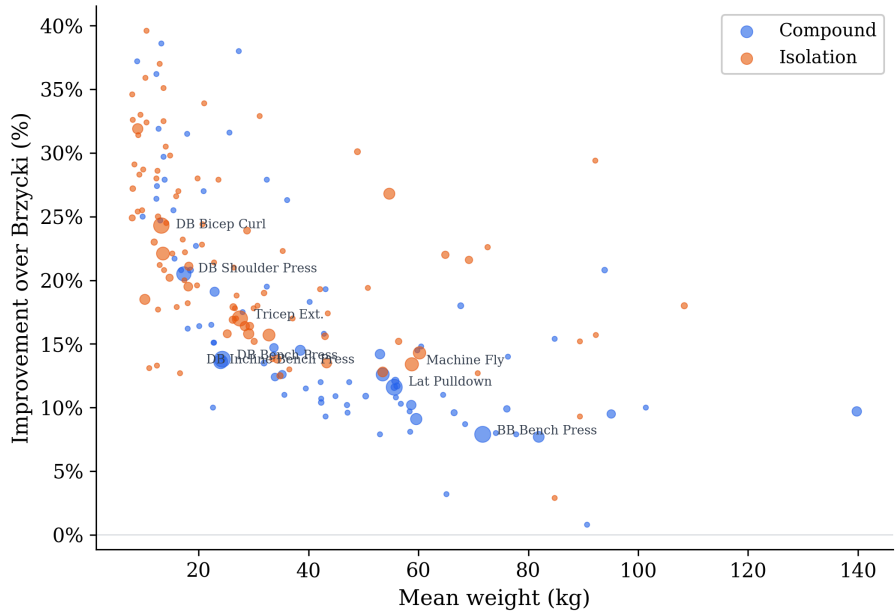

Figure 4. Per-exercise consistency improvement vs. mean weight

Table 8 reports the improvement for the ten most common exercises in the dataset, against all four classical benchmarks.

**Table 8. Consistency improvement for the ten most common exercises.**

| Exercise | Type | Tuples | Mean wt (kg) | vs. Brzycki | vs. Wathen | vs. Epley | vs. Mayhew |
|---|---|---|---|---|---|---|---|
| Barbell Bench Press | Compound | 9,850 | 71.7 | +7.9% | +7.3% | +8.3% | +15.9% |
| Lat Pulldown | Compound | 5,489 | 55.6 | +11.6% | +9.2% | +9.7% | +14.5% |
| Dumbbell Bench Press | Compound | 5,300 | 24.3 | +13.8% | +13.2% | +13.6% | +19.8% |
| Dumbbell Bicep Curl | Isolation | 4,514 | 13.2 | +24.3% | +27.2% | +28.2% | +34.5% |
| Cable Rope Tricep Extension | Isolation | 4,352 | 27.5 | +17.0% | +15.9% | +16.1% | +19.0% |
| Dumbbell Shoulder Press | Compound | 3,859 | 17.3 | +20.5% | +22.7% | +23.4% | +29.4% |
| Dumbbell Incline Bench Press | Compound | 3,645 | 24.0 | +13.6% | +14.7% | +15.2% | +20.8% |
| Machine Fly | Isolation | 3,313 | 58.8 | +13.4% | +8.2% | +8.3% | +10.0% |
| Cable Row | Compound | 3,169 | 53.5 | +12.6% | +10.9% | +11.2% | +15.6% |
| Hammer Curls | Isolation | 3,146 | 13.5 | +22.1% | +24.7% | +25.2% | +30.9% |

Several patterns are visible in Table 8. First, the improvement is positive for every exercise against every benchmark — there is no formula-selection effect where gains against one benchmark come at the expense of another. Second, the relative ranking of exercises by improvement is stable across benchmarks: light isolation exercises (Dumbbell Bicep Curl, Hammer Curls) consistently show the largest gains, while heavy compound exercises (Barbell Bench Press, Cable Row) show the smallest.

Third, the per-exercise rankings shift subtly across benchmarks. For example, the Dumbbell Bicep Curl shows a +24.3% improvement against Brzycki but +34.5% against Mayhew, while the Machine Fly shows +13.4% against Brzycki but only

+10.0% against Mayhew. These differences arise because the classical equations are not uniformly spaced — they differ in how they handle different weight–rep combinations — but in every case the proposed formula outperforms all four.

Even the smallest improvement in the table — +7.3% for the Barbell Bench Press against Wathen — is a meaningful reduction in estimation inconsistency. That the bench press shows the smallest gain is expected: at 71.7 kg average weight, k(w) ≈ 17, where the weight-dependent correction is moderate. It is also the exercise most similar to the conditions under which classical equations were originally derived (Brzycki, 1993; Mayhew et al., 1992).

## Cross-Validation

Table 9 presents the results of the 5-fold user-level cross-validation, with test-set consistency evaluated against all four classical benchmarks.

**Table 9. Five-fold user-level cross-validation results.**

| Fold | a | b | Train SD | Test SD | vs. Brzycki | vs. Wathen | vs. Epley |
| --- | --- | --- | --- | --- | --- | --- | --- |
| 1 | −2.55 | 4.60 | 0.0850 | 0.0838 | +17.7% | +17.1% | +17.5% |
| 2 | −2.50 | 4.58 | 0.0847 | 0.0851 | +17.8% | +17.4% | +17.7% |
| 3 | −2.50 | 4.56 | 0.0852 | 0.0829 | +16.7% | +16.5% | +16.9% |
| 4 | −2.55 | 4.56 | 0.0846 | 0.0853 | +17.5% | +16.8% | +17.2% |
| 5 | −2.55 | 4.58 | 0.0842 | 0.0870 | +18.2% | +17.2% | +17.6% |
| **Mean** | **−2.53** | **4.58** | **0.0847** | **0.0848** | **+17.6%** | **+17.0%** | **+17.4%** |

*Note.* "vs. Brzycki/Wathen/Epley/Mayhew" = reduction in test-set SD(log(1RM)) relative to that benchmark applied to the same held-out users. Overfit = (test SD − train SD) / train SD × 100%.

Three findings are notable. First, the mean improvement on held-out users (+17.6% vs. Brzycki, +17.0% vs. Wathen, +17.4% vs. Epley, +21.8% vs. Mayhew) is virtually identical to the full-sample improvements reported in Table 5, indicating zero effective overfitting. Second, the optimized coefficients are highly stable across folds: a ranges from −2.50 to −2.55 (SD = 0.024), and b ranges from 4.56 to 4.60 (SD = 0.015). Third, the absolute SD(log) values are nearly indistinguishable between train and test (mean: 0.0847 vs. 0.0848), confirming that the formula captures a robust population-level pattern rather than sample-specific noise.

Three of the five folds actually produced *negative* overfit (test SD lower than train SD). This can occur by chance when different user pools happen to have slightly different intrinsic levels of within-tuple consistency, and it underscores that there is no systematic degradation on held-out data.

Figure 5 illustrates the coefficient stability (panel a) and the train–test comparison across folds (panel b). Panel b also plots the Brzycki, Wathen, Epley, and Mayhew test-set baselines, making the gap between classical and proposed consistency visible fold by fold.

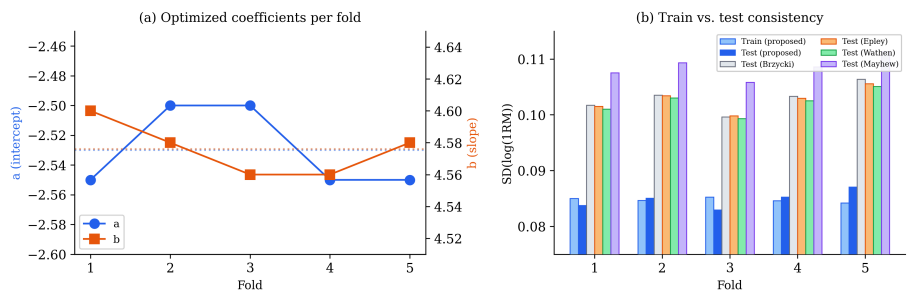

Figure 5. Cross-validation results

## Robustness Analyses

To probe the robustness of the main findings and decompose the sources of improvement, we conducted six supplementary analyses.

## Ablation: Decomposing the Two Innovations

The proposed formula introduces two structural changes relative to Epley: (i) sub-linear repetition scaling (α < 1) and (ii) a weight-dependent conversion factor k(w). An ablation analysis isolates each contribution by testing intermediate models:

- **α-only model:** Uses α = 0.85 with a single optimized fixed k (no weight dependence). Formula: `1RM = w × (1 + (r−1)^0.85 / k)`. Optimized k = 12.55.
- **k(w)-only model:** Uses α = 1.0 (linear, as in Epley) with an optimized weight-dependent k(w). Formula: `1RM = w × (1 + (r−1) / (a + b·ln(w)))`. Optimized a = −4.15, b = 7.60.
- **Full model:** Both innovations. α = 0.85, a = −2.55, b = 4.58.

**Table 10. Ablation: decomposing the two innovations.**

| Model | SD(log(1RM)) | vs. Brzycki |
|---|---|---|
| Brzycki (baseline) | 0.1028 | — |
| α-only (α=0.85, fixed k) | 0.1004 | +2.4% |
| k(w)-only (α=1.0) | 0.0864 | +16.0% |
| Full model | 0.0847 | +17.6% |

The weight-dependent k(w) is the primary driver: it alone accounts for 16.0 of the 17.6 percentage points of improvement (91%). Sub-linear repetition scaling contributes an additional 1.6 percentage points (9%). The synergy between the two innovations — the gap between their sum (18.4%) and the full-model improvement (17.6%) — is minimal, indicating that the two mechanisms operate largely independently.

This decomposition clarifies the paper's main empirical contribution: the finding that the rep-to-1RM conversion factor varies with absolute load, and that modeling this variation drives nearly all the consistency gain.

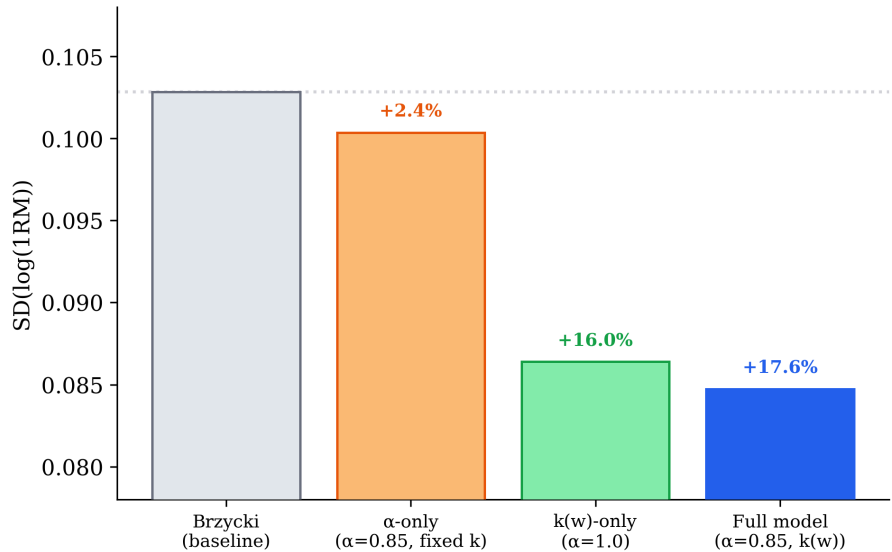

Figure 6. Ablation: decomposing the two innovations

## Sensitivity to α

The rep exponent α was fixed at 0.85 based on preliminary exploration. To assess sensitivity, we reoptimized (a, b) at each α ∈ {0.50, 0.55, ..., 1.00} and evaluated full-sample consistency.

**Table 11. Sensitivity to α (a, b reoptimized at each value).**

| α | Optimized a | Optimized b | SD(log(1RM)) | vs. Brzycki |
|---|---|---|---|---|
| 0.50 | −1.50 | 1.42 | 0.0774 | +24.7% |

| α | Optimized a | Optimized b | SD(log(1RM)) | vs. Brzycki |
|---|---|---|---|---|
| 0.55 | −1.90 | 1.80 | 0.0786 | +23.5% |
| 0.60 | −2.00 | 2.12 | 0.0798 | +22.4% |
| 0.65 | −2.45 | 2.58 | 0.0809 | +21.3% |
| 0.70 | −2.85 | 3.10 | 0.0820 | +20.2% |
| 0.75 | −3.45 | 3.74 | 0.0831 | +19.2% |
| 0.80 | −4.20 | 4.54 | 0.0841 | +18.2% |
| **0.85** | **−2.55** | **4.58** | **0.0847** | **+17.6%** |
| 0.90 | −3.00 | 5.44 | 0.0853 | +17.0% |
| 0.95 | −3.55 | 6.46 | 0.0859 | +16.5% |
| 1.00 | −4.15 | 7.60 | 0.0864 | +16.0% |

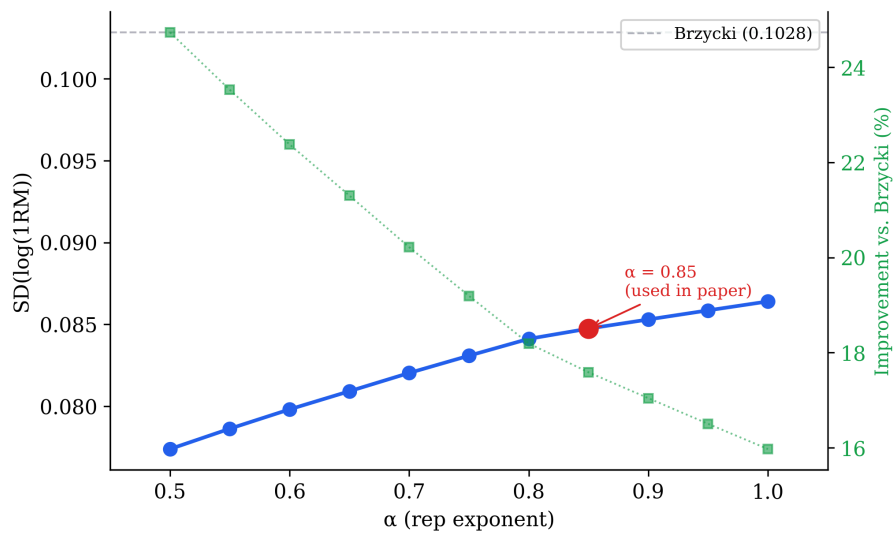

Figure 7. Sensitivity to α: reoptimized (a, b) at each α

Two patterns emerge. First, the improvement over Brzycki is positive across the entire range of α, from +16.0% (α = 1.0, equivalent to adding only the weight-dependent k(w) to Epley) to +24.7% (α = 0.50). The paper's central finding — that a weight-dependent conversion factor substantially improves consistency — holds regardless of the chosen α.

Second, the relationship between α and SD is monotonically increasing: lower α values yield lower (better) within-tuple SD. The improvement at α = 0.50 substantially exceeds that at α = 0.85. We interpret this monotonic pattern with caution: lower α values compress the rep-count term, making the formula less sensitive to differences between, say, 10 and 15 reps. This reduced sensitivity mechanically lowers within-tuple variance. Whether it reflects a genuinely better model of the rep–1RM relationship or simply a less informative one cannot be determined from the consistency metric alone — that distinction requires external validation against directly measured 1RM values. We retain α = 0.85 as a moderately conservative choice that avoids excessive compression of the rep term while still providing meaningful improvement.

### Tuple-Size Stratification

Because 81.3% of tuples contain exactly 2 sets, a potential concern is that the results are driven by 2-set tuples, whose within-tuple SD is computed from only two observations and may be noisy. Table 12 stratifies by tuple size.

**Table 12. Consistency by tuple size.**

| Tuple size | n tuples | % of total | SD(log) Brzycki | SD(log) Ours | Improvement |
|---|---|---|---|---|---|
| n = 2 | 110,298 | 81.3% | 0.1033 | 0.0850 | +17.8% |

| Tuple size | n tuples | % of total | SD(log) Brzycki | SD(log) Ours | Improvement |
|---|---|---|---|---|---|
| n ≥ 3 | 25,432 | 18.7% | 0.1007 | 0.0838 | +16.8% |

The improvement is consistent across tuple sizes: +17.8% for 2-set tuples and +16.8% for tuples with 3 or more sets. The slightly larger improvement for 2-set tuples is expected (the SD estimate is noisier with only 2 observations, giving more room for any formula to improve), but the magnitude of the difference is small (1 percentage point). The core finding is not an artifact of 2-set tuples.

### Window-Size Sensitivity

The 14-day window used to define tuples was chosen as a balance between stability (short windows) and data accumulation (long windows). Table 13 shows results for 7-day and 28-day windows.

**Table 13. Consistency by window size.**

| Window (days) | n tuples | n sets | SD(log) Brzycki | SD(log) Ours | Improvement |
|---|---|---|---|---|---|
| 7 | 45,804 | 94,777 | 0.1034 | 0.0846 | +18.2% |
| 14 (primary) | 135,730 | 303,494 | 0.1028 | 0.0847 | +17.6% |
| 28 | 128,406 | 303,494 | 0.1043 | 0.0859 | +17.6% |

The improvement is highly stable across window sizes: +18.2% at 7 days, +17.6% at 14 days, and +17.6% at 28 days. The 7-day window yields fewer qualifying tuples (shorter periods accumulate less variation), while the 28-day window shows slightly higher baseline SD (true strength may change over four weeks), but the proposed formula's advantage is unchanged.

### k(w) Floor Activation

The computational guard k(w) ≥ 0.5 was imposed to prevent division by near-zero values. Out of 303,494 sets, only 173 (0.06%) had unguarded k(w) < 0.5 — all involving weights below 2 kg. The guard is effectively inert for the analysis.

The formula produces k(w) = 0 at w = 1.75 kg and k(w) = 0.5 at w = 1.95 kg. For the lightest exercise with meaningful data (Cable Lateral Raise, mean weight 7.9 kg, k(w) = 6.9), the floor is never approached. This confirms that the negative intercept in k(w) does not create numerical instabilities within the range of weights observed in practice.

### Equipment-Type Stratification

To assess whether the improvement varies by equipment category — and whether the weight-dependence of k(w) is confounded by equipment-specific logging conventions — we stratified exercises by equipment type inferred from exercise names.

**Table 14. Consistency improvement by equipment type.**

| Equipment | n exercises | n tuples | Mean weight (kg) | SD(log) Brzycki | SD(log) Ours | Improvement |
|---|---|---|---|---|---|---|
| Barbell | 81 | 27,583 | 61.8 | 0.0832 | 0.0730 | +12.3% |
| Cable | 50 | 19,664 | 30.2 | 0.1196 | 0.0970 | +18.9% |
| Dumbbell | 99 | 35,151 | 17.3 | 0.1053 | 0.0820 | +22.1% |
| Machine | 43 | 25,371 | 51.5 | 0.1081 | 0.0908 | +16.0% |
| Other | 115 | 27,961 | 49.7 | 0.1025 | 0.0857 | +16.4% |

The improvement is positive for every equipment category, ranging from +12.3% (Barbell) to +22.1% (Dumbbell). The pattern mirrors the weight dependence observed in the per-exercise analysis: dumbbell exercises (lowest mean weight) show the largest improvement, barbell exercises (highest mean weight) the smallest. This is consistent with the k(w) mechanism operating through actual load magnitude rather than through an artifact of equipment-specific logging conventions — if logging conventions were the primary driver, we would expect machine exercises (where "weight" refers to a stack setting that is not directly comparable across manufacturers) to behave differently from cable exercises, but both show improvement squarely in the range predicted by their mean weights.

# Discussion

## Summary of Findings

This study is, to our knowledge, the first to derive a 1RM prediction equation from large-scale, real-world resistance training data. The underlying dataset — drawn from Fitbod, a consumer fitness app with over a million users — contained 37.7 million logged sets from approximately 66,000 users across 505 exercises. The analysis sample used for optimization contained 303,494 near-failure sets from 14,966 users across 388 exercises spanning 16 muscle groups. Nuzzo et al. (2024), the most comprehensive meta-analysis to date, aggregated 269 studies encompassing 7,289 individuals — roughly half the number of users in our sample alone.

The core methodological contribution is the internal consistency criterion: a way to evaluate and compare 1RM prediction formulas using observational training data, without requiring laboratory-based maximal testing. By measuring how consistently a formula maps different weight–rep combinations to the same estimated 1RM for the same person on the same exercise within a short time window, we can identify formulas that better capture the underlying structure of the rep–weight–1RM relationship — regardless of whether the user was training to true failure or has ever performed a formal 1RM test.

Using this criterion, we optimized a two-parameter generalization of the classical Epley equation in which the rep-to-1RM conversion factor is allowed to vary logarithmically with the weight lifted:

```
1RM = w × (1 + (r − 1)^0.85 / (−2.55 + 4.58 × ln(w)))
```

This formula reduced within-tuple inconsistency by 17–22% relative to all four classical benchmarks tested (Brzycki, Epley, Wathen, and Mayhew). The improvement was positive for every one of the 183 exercises with sufficient data, ranging from +1% for heavy barbell lifts to +40% for light dumbbell and cable movements. Five-fold user-level cross-validation — the most stringent form, where the formula must generalize to entirely new users it has never seen — confirmed near-zero overfitting: the mean test-set improvement was indistinguishable from the full-sample improvement across all four benchmarks.

The ablation analysis reveals a clear hierarchy: the weight-dependent k(w) accounts for 91% of the improvement over Brzycki, while the sub-linear rep exponent (α = 0.85) contributes the remaining 9%. The central finding is that the relationship between repetitions and percentage of 1RM depends on the absolute load. At lighter weights (typically isolation exercises involving smaller muscle groups), each additional repetition carries more information about maximal capacity than at heavier weights (typically compound movements involving larger muscle groups). Classical equations, by using a single conversion factor regardless of load, systematically underestimate the "value" of each rep for light exercises and slightly overestimate it for heavy ones. The weight-dependent k(w) function corrects this mismatch — and the fact that it improves consistency for all 183 tested exercises, with no exceptions, suggests the effect is genuine and pervasive.

This finding is consistent with, and substantially extends, prior work. Shimano et al. (2006) observed that repetition capacity at a given %1RM differs across exercises in small laboratory samples. Nuzzo et al. (2024) identified exercise type as the dominant moderator of the rep–%1RM relationship in their meta-regression. Our results quantify this moderation for the first time at scale, across hundreds of exercises, and embed it directly into a prediction formula.

## Practical Implications

**For coaches and practitioners.** The formula is a simple closed-form expression that can be computed by hand or in a spreadsheet. For a 13 kg dumbbell curl at 10 reps, the classical Brzycki equation estimates a 1RM of about 17 kg; the proposed formula estimates about 22 kg. For a 100 kg bench press at 5 reps, the two formulas agree within 2 kg. Coaches working with athletes across a variety of exercises — not just the barbell lifts for which classical equations were originally calibrated — may find the proposed formula provides more stable day-to-day tracking of estimated maxes, especially for accessory and isolation work.

**For fitness apps and software.** Consumer fitness apps that track estimated 1RM over time currently rely on classical equations derived from small bench-press studies in the 1980s and 1990s. These equations are applied, without modification, to hundreds of exercises they were never designed for — dumbbell lateral raises, cable tricep extensions, machine flys — because no better alternative existed. The proposed formula fills this gap. It requires no additional data collection, no per-exercise calibration, and no change to the user experience; it is a single equation that can replace the classical one in any codebase. And the improvement is largest for exactly the exercises that apps most need to get right: the diverse array of dumbbell, cable, and machine movements that make up the bulk of a typical user's training program.

**For researchers.** The internal consistency criterion offers a scalable methodology for developing and evaluating 1RM prediction formulas on observational training data. This opens formula development to datasets orders of magnitude larger than what laboratory testing can produce — and, critically, to exercise diversity that no lab study could practically achieve. A laboratory validation study testing 10 exercises on 100 participants requires 1,000 maximal tests; our approach evaluated 388 exercises on nearly 15,000 users with no maximal testing at all. The two methodologies are complementary (see below), but the scalability of the observational approach makes it uniquely suited to questions that require broad exercise coverage.

## Methodological Considerations

The strengths and limitations of this study are best understood in contrast with the laboratory paradigm that has dominated 1RM prediction research.

**Internal vs. external validation.** Laboratory studies validate prediction equations by comparing estimated 1RM to a directly measured one-rep max — the gold standard for *absolute accuracy*. Our internal consistency criterion does not measure absolute accuracy; a formula that systematically overestimates 1RM by 10% for everyone would still score well. The 17–22% improvement should therefore be interpreted as evidence that the proposed formula better captures the *relative* structure of the rep–weight–1RM relationship, not as a direct claim about absolute prediction error.

However, the converse limitation is equally important and less often acknowledged: laboratory validation tells you how well a formula works for the specific exercises and populations tested — typically the bench press, sometimes the squat or leg press, performed by young trained males under controlled conditions. Whether those results generalize to a 55-year-old woman performing cable rows in a commercial gym is an open question that lab studies, by design, cannot answer. Our dataset captures precisely this kind of diversity: 388 exercises, ages 18–82, both genders, naturalistic training conditions with self-selected loads, rest intervals, and effort levels. If a prediction formula is going to be used broadly — as it is in every fitness app on the market — it should be evaluated broadly.

The ideal validation, which we hope future work will pursue, would combine both approaches: laboratory-tested 1RM data across a range of exercises and populations, analyzed alongside the kind of large-scale observational consistency data presented here.

**Proximity to failure.** Despite our filtering for AMRAP-flagged and fatigue-detected sets, we cannot be certain that every set in the sample was performed at or near momentary muscular failure. If some sets had substantial repetitions in reserve (Steele et al., 2017), the 1RM estimates derived from them would be systematically biased downward. Importantly, this bias is identical across all formulas and does not affect the comparative finding. It does mean that the formula's coefficients reflect the behavior of near-failure training sets — which is also the context in which the formula is most likely to be applied.

**Population representativeness.** The sample is predominantly male (80%) and concentrated in the 25–39 age range, reflecting the user base of a consumer fitness app. This is a meaningful limitation, but it is worth noting that this population is substantially more diverse than the samples used to derive or validate classical equations. Brzycki (1993) and Epley (1985) published no sample at all; Mayhew et al. (1992) tested 435 college students on the bench press; the validation studies of LeSuer et al. (1997), Reynolds et al. (2006), and Mayhew et al. (2008) had sample sizes of 67, 70, and 103, respectively — all drawn from narrow demographic pools. The present dataset, while imperfect, is the most diverse sample on which any 1RM prediction equation has been evaluated.

**The role of α.** The sensitivity analysis revealed that the consistency metric improves monotonically as α decreases from 1.0 to 0.5. This means our choice of α = 0.85, while producing a substantial improvement over classical equations, is not the metric-optimal value. We retained α = 0.85 for two reasons. First, the ablation analysis demonstrates that α contributes only 9% of the total improvement — the weight-dependent k(w) is the dominant innovation regardless of α. Second, very low α values (e.g., 0.50) heavily compress the rep term, making the formula less sensitive to differences in repetition count. Whether this compression better models the true rep–1RM relationship or simply reduces measurement noise cannot be determined without external validation. An α of 0.85 represents a conservative middle ground that preserves meaningful rep-sensitivity while still capturing the sub-linear pattern evident in the data.

**Single set of coefficients across all exercises.** The proposed formula uses weight as a proxy for exercise type: lighter exercises tend to involve smaller muscle groups, and heavier exercises larger ones. This is a useful approximation but an imperfect one. Two exercises at the same weight — say, a 25 kg dumbbell bench press and a 25 kg concentration curl — may have genuinely different rep–%1RM curves that the formula cannot distinguish. An exercise-specific or muscle-group-specific model would likely perform better, and the dataset is large enough to support such modeling.

**Fixed α and computational constraints.** The analysis used a 5% user sample for computational tractability. While the optimized coefficients were stable across subsample sizes, joint optimization of all three parameters on the full dataset — roughly 750 million sets from over 1.3 million users — could potentially refine the results.

## Future Research Directions

**External validation.** One interesting next step is to test the formula against directly measured 1RM values across a range of exercises and populations. A study collecting multi-rep sets and a true 1RM test on the same exercise and day — for exercises spanning a wide weight range — would provide a definitive test of whether the internal consistency improvement translates to improved absolute accuracy.

**Exercise-specific models.** The current formula uses weight as an indirect proxy for the physiological factors that modulate the rep–%1RM curve. A more principled approach would allow the formula's parameters to vary by exercise, muscle group, or exercise category. The present dataset is large enough to support such estimation, and hierarchical Bayesian modeling would be a natural framework — allowing exercise-specific coefficients to borrow strength from the overall population estimate.

**Joint optimization of α.** The sensitivity analysis suggests that lower α values may yield better consistency. A natural extension would be to jointly optimize α, a, and b — either by extending the grid search to three dimensions (computationally expensive but feasible on the full dataset) or by using gradient-based optimization. Cross-validating α within each fold would also address whether the metric improvement at lower α generalizes to held-out users.

**Incorporating fatigue and set order.** Our filtering strategy — retaining only the first near-failure set per exercise per workout — discards a large amount of data. A formula that explicitly models within-session fatigue could potentially use all available sets and provide more precise estimates. No existing 1RM prediction equation accounts for set order, and the data to support such modeling now exist.

**Interaction with training variables.** Nuzzo et al. (2024) found that sex, age, and training status had minimal moderating effects on the rep–%1RM relationship, but this has not been tested at the scale of data available here. Investigating whether the optimal k(w) varies by demographic subgroup would be of both scientific and practical interest.

**Combination with velocity data.** As barbell and dumbbell velocity sensors become more common, it may be possible to combine repetition-based and velocity-based 1RM estimates into a hybrid model. The weight-dependent structure proposed here could serve as a prior for individualized velocity–load profiles, bridging the gap between population-level prediction equations and the fully individualized approach that the velocity-based literature has shown to be superior (Greig et al., 2023).

---